\documentclass[runningheads]{llncs}
\usepackage{graphicx,amssymb,booktabs}
\usepackage{graphicx}
\usepackage{bm}
\usepackage{array}
\usepackage{amsmath}
\usepackage{float}
\usepackage{bbding}
\usepackage[doipre={doi:~}]{uri}
\begin{document}
	\title{CLCI-Net: Cross-Level fusion and Context Inference Networks for Lesion Segmentation of Chronic Stroke}
	\titlerunning{CLCI-Net for Lesion Segmentation of Chronic Stroke}
	\renewcommand{\thefootnote}{\fnsymbol{footnote}} 
	\author{Hao Yang\inst{*1,2}\and
		Weijian Huang\inst{*3}\and
		Kehan Qi\inst{1}\and
		Cheng Li\inst{1}\and
		Xinfeng Liu\inst{4}\and
		Meiyun Wang\inst{5}\and
		Hairong Zheng\inst{1}\and
		Shanshan Wang\inst{1 }{\Envelope}}
	\authorrunning{H. Yang et al.}
	\institute{Paul C. Lauterbur Research Center for Biomedical Imaging, Shenzhen Institutes of Advanced Technology, Chinese Academy of Sciences, Shenzhen, China \\  \email{sophiasswang@hotmail.com}\and
		University of Chinese Academy of Sciences, Beijing, China\and
		School of Biomedical Engineering, Health Science Center, Shenzhen University, Shenzhen, China\and
		Guizhou Provincial People's Hospital, Guizhou, China  \and
		Department of Radiology, Henan Provincial People’s Hospital, Henan, China}
	\footnotetext{* These authors contruibuted equally to this work.}
	\maketitle              % typeset the header of the contribution
	\begin{abstract}
		Segmenting stroke lesions from T1-weighted MR images is of great value for large-scale stroke rehabilitation neuroimaging analyses. Nevertheless, there are great challenges with this task, such as large range of stroke lesion scales and the tissue intensity similarity. The famous encoder-decoder convolutional neural network, which although has made great achievements in medical image segmentation areas, may fail to address these challenges due to the insufficient uses of multi-scale features and context information. To address these challenges, this paper proposes a Cross-Level fusion and Context Inference Network (CLCI-Net) for the chronic stroke lesion segmentation from T1-weighted MR images. Specifically, a Cross-Level feature Fusion (CLF) strategy was developed to make full use of different scale features across different levels; Extending Atrous Spatial Pyramid Pooling (ASPP) with CLF, we have enriched multi-scale features to handle the different lesion sizes; In addition, convolutional long short-term memory (ConvLSTM) is employed to infer context information and thus capture fine structures to address the intensity similarity issue. The proposed approach was evaluated on an open-source dataset, the Anatomical Tracings of Lesions After Stroke (ATLAS) with the results showing that our network outperforms five state-of-the-art methods. We make our code and models available at \url{https://github.com/YH0517/CLCI_Net}.
		\keywords{Deep learning  \and Chronic stroke \and Segmentation \and Cross-Level}
	\end{abstract}
	\section{Introduction}
	Clinical intervention is necessary for the treatment and prognosis of patients with chronic stroke. Currently, high-resolution T1-weighted (T1W) anatomical magnetic resonance imaging is commonly used to understand the relationship between brain behavior and recovery after stroke in clinics. Quantifying and evaluating a patient's condition requires manually mapping the lesion area in clinical work. This is a time-consuming, labor-intensive, and subjective process \cite{[1]}. Therefore, there is a need for a reliable method that can help doctors automatically identify areas of the lesion. 
	
	With the rapid development of deep learning, Convolutional Neural Networks (CNNs) have shown a great potential in medical image analysis in recent years. In particular, U-Net \cite{[2]}, which adopts the encoder–decoder structure and skip-connections to combine contextual information, has achieved great success in medical image segmentation tasks. However, the local receptive field and the efficiency of feature re-use are limited by the fixed convolution size and single downsampling path in U-Net, that may not be conducive to deal with the problems of the great variation in size and boundary ambiguity of lesions in stroke segmentation. Atrous Spatial Pyramid Pooling (ASPP) \cite{[3]} is proposed for the fusion of multi-scale features. This structure combines the features, which are generated by several parallel dilated convolutions with different dilated rates, to form multi-scale predictions. However, this multi-scale feature fusion strategy is only performed at the same sampling level regardless of the scale of different downsampling levels. Meanwhile, U-Net assembles a more elaborate prediction in the decoding phase by connecting context features before going through a series of convolution operations. Although this approach allows the network to perform local estimation through global guidance, direct stacking of the convolution channel may not be sufficient for the fusion of different levels of information. Recurrent Neural Network (RNN) can improve the semantic segmentation result by inferring the global context. Yang et al. Proposed a multi-directional RNN encodes spatial sequentiality to combat boundary blur for significant refinement \cite{[4]}; Li et al. takes pyramidal features to refine the segmentation mask progressively \cite{[5]}. However, the above-mentioned work treated RNN as a post-processing method to refine initial segmentation result. Introducing the RNN to the decoding phase might be a more effective approach. 
	
	This paper presents a new end-to-end neural network framework, Cross-Level fusion and Context Inference Network (CLCI-Net), to address the challenges of chronic stroke segmentation in T1 images. During the encoding phase, we improved the way the information flows. Unlike ResNet \cite{[6]} and DenseNet \cite{[7]}, information in different downsampling stages is stacked to exploit the potential of cross-level information (high-level semantics and low-level textures) to complement each other. We also used this strategy to expand the ASPP structure and better deal with the problem of large differences in the location, shape and size of stroke lesions by integrating more scale information. To improve the integrity of the lesion prediction results, we replaced the direct stacking of spatial during the decoding phase by inferring the context information using convolutional LSTM (ConvLSTM) \cite{[8]} to improve the integrity of the model prediction. The main contributions include the development of a new CLCI-Net, which has the following innovations: a Cross-Level feature Fusion (CLF) strategy is developed to achieve smoother information flow and thus facilitate more sufficient utilization of extracted features; multi-scale information is enriched to handle the different lesion sizes by integrating CLF with ASPP; Last but not least, convolutional long short-term memory (ConvLSTM) is employed to infer context information and thus capture feature details to address the intensity similarity issue. The proposed model’s effectiveness was evaluated on an open-source dataset, the Anatomical Tracings of Lesions After Stroke (ATLAS) with the results showing that our network outperforms five state-of-the-art methods.
	
	\section{Method}
	\begin{figure}
		\includegraphics[width=\textwidth]{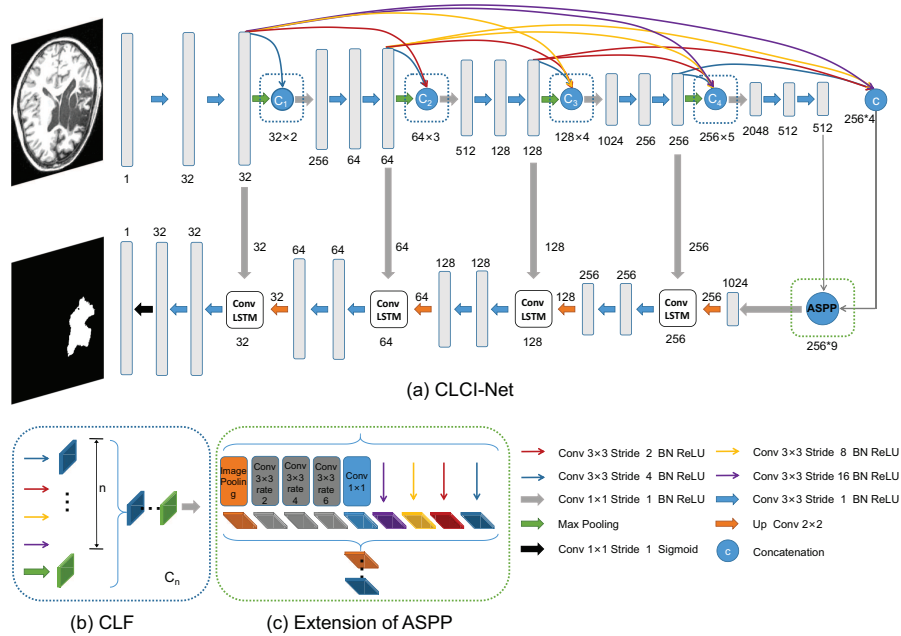}
		\caption{(a) The architecture of the proposed CLCI-Net. More robust and finer segmentation results are obtained by multi-scale feature fusion and context inference. (b) CLF strategy. Enhanced gradient propagation and feature re-use by connecting features across the downsampling level. (c) Extension of ASPP.} \label{fig1}
	\end{figure}
	
	\subsubsection{Cross-Level Information}
	
	The low-level layers of a neural network tend to extract image texture features, and more semantic information is encoded along with the increase of network depth. There have been many experiments showing that deeper neural networks bring better performance. But the network with too many layers may encounter problems such as vanishing/exploding gradient. ResNet uses shortcut connections to skip one or more layers and add their outputs, effectively alleviating the above issues. However, only a single connection between different levels may not fully re-use the features. DenseNet extends the concept of concatenation between adjacent levels so that the input of each layer comes from all of the previous feature maps. However, this connection strategies is only used in the same downsampling level, thus lacking the information complementing ability between different downsampling levels. Based on this, we propose a CLF strategy in which the output of each downsampling layer is aggregated with all of the previous features before the downsampling operation. This strategy allows the integration of features from different sampling levels to enhance connection and complementarity between cross-level information. It is worth noting that for different feature aggregations, convolutions with different strides are used to ensure a consistent resolution between the features, as shown in Fig.~\ref{fig1}(a).	
	\subsubsection{Multi-scale Feature}
	Multi-scale features are an important factor to improve the segmentation performance. Chen et al. proposed ASPP \cite{[3]}, which integrates the features from different receptive fields through multiple parallel-distributed dilated convolutions, and obtaining more refined and robust features. In this paper, we extend the ASPP module using CLF, as shown in Fig.~\ref{fig1}(c). Specifically, the four levels of features from the downsampling path are combined with the five scale features in the original ASPP to produce features of nine scales. Thereby, the network not only obtains five scale features from high-level semantic information, but also gets the texture and position information from CLF in decoding.
	
	\subsubsection{Context Inference}
	Due to the U-shaped structure, U-Net \cite{[2]} has achieved great success in the medical segmentation task. The skip-connections combine the high-resolution features from the contracted path with the upsampled outputs, allowing the network to perform local estimation under global guidance \cite{[2],[9]}.\begin{figure}
		\includegraphics[width=\textwidth]{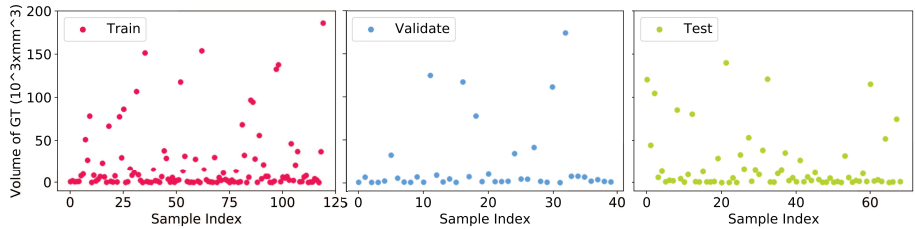}
		\caption{The distribution of stroke lesions size in three sets, in which the proportion of large lesions and small lesions in each set is relatively balanced.} \label{fig2}
	\end{figure} We inherited this contextual information fusion strategy. However, simple feature concatenation may not be able to fully recover the lost information due to downsampling. RNN has the ability to model the global context and improve semantic segmentation by associating pixel-level and local information. Inferring from the context information, the ConvLSTM replicates the true value of its state and accumulates external signals in the sequence step, enhancing local prediction \cite{[8]}.    
	\begin{figure}
		\includegraphics[width=\textwidth]{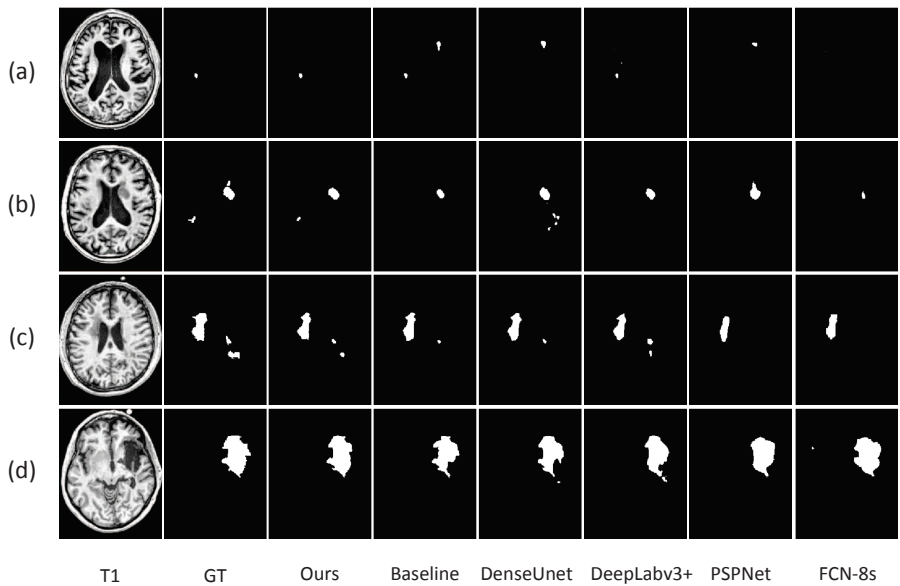}
		\caption{Comparisons of our method, Baseline, DenseUnet, DeepLabv3+, PSPNet, and FCN-8s on four different patients.} \label{fig3}
	\end{figure}
	\subsubsection{Cross-Level Fusion and Context Inference Network}
	Our proposed CLCI-Net is shown in Fig.~\ref{fig1}(a). We mainly use convolution kernels with size of $3 \times 3$ and $1 \times 1$. All the convolution layers are followed by batch normalization and ReLU activation. The feature numbers of different layers are listed in the figure. 
	
	In the encoding phase, we propose a CLF strategy to increase the efficiency of feature map reuse and to fuse feature information across the downsampling level. Specifcally, as shown in Fig.~\ref{fig1}(b), we use CLF strategy in four downsampling layers and an ASPP to ensure enhanced cross-level feature connection and complementarity between cross-level information. As shown in Fig.~\ref{fig1}(c), further extending ASPP with CLF strategy enables the model to benefit from the multi-scale transformation of high-level semantic information and low-level information of position and texture. In the decoding phase, we replaced the traditional concatenation operation with ConvLSTM to capture more fine-grained structure loss by inferring context information. Finally, a 1x1 convolution followed by Sigmoid activation is adopted to output a feature probability map that is consistent with the original image size.
	\begin{figure}
		\includegraphics[width=\textwidth]{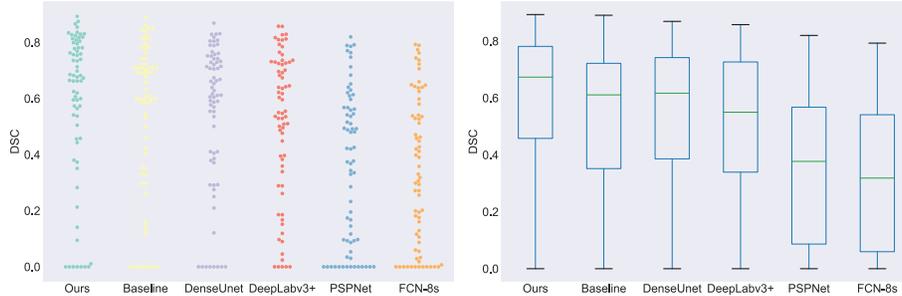}
		\caption{The DSC Statistical Diagram of each model. The left figure displays the DSC score distribution of each sample as a beeswarm. The figure on the right shows the corresponding box plot to show the overall DSC statistics.} \label{fig4}
	\end{figure}
	
	\section{Experiments and Results}
	
	\subsubsection{Experiments}
	We adopted the subset of open source dataset ATLAS, which contains 229 subjects. We randomly selected 120 subjects for training, 40 for validation, and 60 for testing. Since the size of stroke lesion has a great influence on network performance, we have calculated the distribution of lesion sizes in the three groups. As shown in Fig.~\ref{fig2}, the proportions of large lesions and small lesions in the three groups are roughly balanced. In addition, we croped the images from $233 \times 197$ to $224 \times 176$ in order to adapt to the input size of the network. 
	
	We have compared our approach with different outstanding methods, including DenseUnet, DeepLabv3+, PSPNet, and FCN-8s \cite{[10]}. Specifically, we adjusted the parameter size of U-Net, and added the BN layer after each convolution layer, to improve network performance and accelerate convergence. This method was chosen as a comparison baseline. The training parameters of our approach were set as follows: used Gaussian function to initialize the weight, use Dice Loss as the loss function, and the Adam optimizer for gradient optimization. The learning rate is set to 0.0001.

	\subsubsection{Qualitative Results}
	
	Several challenging cases are shown in Fig.~\ref{fig3}. We can see our model segmentation results are consistent better than other outstanding methods. Specifically, (a) shows that Baseline, DenseUnet, and PSPNet \cite{[11]} incorrectly identified the tissue with low-intensity signals as the stroke lesion area, while our model presents more accurate segmentation. Furthermore, for difficult small lesion samples as shown in (a-c), our methods presented stronger capability in identifying and segmenting them. Last but not least, for the large lesions shown in (d), our model can get more detailed boundary information. This proves the effectiveness of our proposed method in improving the segmentation accuracy.
	
	\subsubsection{Quantitative Results}
	In this section, we demonstrate the superiority of our proposed method through calculate indicators: Dice Similarity Coefficient (DSC), Precision, Recall, Volumetric Overlap Error (VOE), and Relative Volume Difference (RVD).
	\begin{table}[tp]  
		
		\centering  
		\fontsize{10}{12}\selectfont  
		\caption{Comparison of proposed method with several popular segmentation frameworks.}
		  
		\label{tab1}
		\begin{tabular}{c|ccccc}  
			\toprule  		
			\bf Method&\bf \ \ DSC\ \ &\bf Precision&\bf \ \ Recall\ \ &\bf \ \ VOE\ \ \bf&\bf \ \ RVD\ \ \cr 
			\midrule  
			FCN-8s&0.337&0.485&0.334&76.5&19.6\cr  
			PSPNet&0.375&0.502&0.361&73.5&\bf13.6\cr  
			DeepLabv3+&0.507&0.586&0.527&62.1&32.5\cr  
			DenseUnet&0.543&0.614&0.553&58.8&25.6\cr  
			Baseline&0.54&0.632&0.544&58.7&31.7\cr 
			Ours&\bf0.581&\bf0.649&\bf0.581&\bf54.6&25.4\cr  
			\bottomrule  
		\end{tabular}  
	\end{table}
	In Table~\ref{tab1}, we quantitatively compare our method to some of the currently widely used segmentation methods. We can observe that our model has the highest scores on the main indicator (DSC) as well as the auxiliary indicators. The DSC of our model is 3.8\% higher than that of DenseUnet, which is the second-best method. This demonstrates that our model could achieve promising segmentation performance.
	
	We show the DSC statistical distribution plots for different models in Fig.~\ref{fig4}. The image on the left shows the specific distribution of DSC scores. It can be seen that our method has a denser distribution at high DSC values than others, which is confirmed by the boxplot on the right. This illustrates that our model has superior segmentation performance on the overall data, not limited to individual samples.
	
	To investigate the contribution of each component to the proposed framework, we list the various combinations in Table~\ref{tab2}. The original ASPP shows a limited improvement in model performance, which may be due to the insufficient multi-scale information extraction of features. CLF improves DSC by 1.1\%, and achieves the optimal precision and RVD, indicating that CLF can help the model to regulate the details and make the output more detailed. The inference structure improves DSC by 1.8\% and generates the optimal recall, which enhances the ability to capture features by inferring context information. In addition, we also compared various combinations. All results show that each structure has a certain improvement relative to the baseline. This further proves that our proposed scheme can improve the performance of existing models. 
	\begin{table}[tp]  
		
		\centering  
		\fontsize{10}{12}\selectfont  
		
		\caption{Verify the effects of each component on the Baseline. Among them, the ASPP: ASPP operation used in DeepLabv3+. CLF: Our proposed cross-level connection strategy. Inference: Infer context information using ConvLSTM.}	\label{tab2}  
		\begin{tabular}{ccc|ccccc}  
			\toprule  
			
			\bf \ ASPP\ &\bf \ CLF\ &\bf \ Inference\ &\bf \ \ DSC\ \ &\bf Precision&\bf \ \ Recall\ \ &\bf \ \ VOE\ \ \bf&\bf \ \ RVD\ \ \cr  
			\midrule  
			&&&0.54&0.632&0.544&58.7&31.7\cr  
			\checkmark&&&0.546&0.64&0.537&57.7&31\cr  
			 &\checkmark& &0.551&\bf0.66&0.535&57.7&\bf12.8\cr  
			 & &\checkmark&0.558&0.603&\bf0.601&57&53.4\cr  
			\checkmark&\checkmark& &0.559&0.65&0.553&56.5&26.2\cr  
			\checkmark& &\checkmark&0.567&0.622&0.572&55.8&43.6\cr  
			&\checkmark&\checkmark&0.568&0.599&0.597&55.7&37.9\cr  
			\checkmark&\checkmark&\checkmark&\bf0.581&0.649&0.581&\bf54.6&25.4\cr
			\bottomrule  
		\end{tabular}  
	\end{table}

	\section{Discussion and Conclusion}
	
We propose a new approach CLCI-Net to automatically segment chronic stroke lesions from T1 weighted MR images. CLCI-Net is novel in three aspects 1) a new CLF strategy is developed to make full use of different levels of features, which also has the merit of avoiding gradient explosion/vanishing and therefore facilitates deep feature extraction and utilization; 2) CLF is further employed to extend ASPP to address the challenges with the big variety of lesion scales. 3) ConvLSTM has been adopted to replace the commonly used spatial stacking operation, with more fine structures captured to distinguish different but visually similar tissues; The proposed approach has been evaluated on a famous open dataset ATLAS and compared to five state-of-the-art methods. Experimental results show that the proposed CLCI-Net has obtained the best performance and presented greatest robustness to the large range of stroke lesion scales and the tissue intensity similarity. 

\subsubsection{Acknowledgments}
This research was partly supported by the National Natural Science Foundation of China (61601450, 61871371, 81830056), Science and Technology Planning Project of Guangdong Province (2017B020227012, 2018B01 0109009), the Basic Research Program of Shenzhen (JCYJ20180507182400762), Youth Innovation Promotion Association Program of Chinese Academy of Sciences (2019351).

	%\bibliographystyle{splncs04}
	%\bibliography{mybibliography}
	%
	
\end{document}